\documentclass[conference]{IEEEtran}
\IEEEoverridecommandlockouts
\usepackage{cite}
\usepackage{amsmath,amssymb,amsfonts}
\usepackage{algorithmic}
\usepackage{graphicx}
\usepackage{textcomp}
\usepackage{xcolor}
\usepackage{hyperref}
\usepackage{url}
\def\BibTeX{{\rm B\kern-.05em{\sc i\kern-.025em b}\kern-.08em
    T\kern-.1667em\lower.7ex\hbox{E}\kern-.125emX}}

\begin{document}

\title{quEEGNet: Quantum AI for Biosignal Processing}

\author{
\IEEEauthorblockN{Toshiaki Koike-Akino, Ye Wang}
\IEEEauthorblockA{\textit{Mitsubishi Electric Research Laboratories (MERL)} \\
201 Broadway, Cambridge, MA 02139, USA \\
\{koike, yewang\}@merl.com}
}
\maketitle

\begin{abstract}
In this paper, we introduce an emerging quantum machine learning (QML) framework to assist classical deep learning methods for biosignal processing applications.
Specifically, we propose a hybrid quantum-classical neural network model that integrates a variational quantum circuit (VQC) into a deep neural network (DNN) for electroencephalogram (EEG), electromyogram (EMG), and electrocorticogram (ECoG) analysis.
We demonstrate that the proposed quantum neural network (QNN) achieves state-of-the-art performance while the number of trainable parameters is kept small for VQC.
\end{abstract}


\section{Introduction}

The great advancement of artificial intelligence (AI) techniques based on deep neural networks (DNN) has enabled practical development of human-machine interfaces (HMI) including brain-computer interfaces (BCI) through the analysis of the user’s physiological data~\cite{faust2018deep}, such as electroencephalogram (EEG)~\cite{lawhern2018eegnet} and electromyogram (EMG)~\cite{atzori2016deep}. 
However, such biosignals are highly prone to variation depending on the biological states of each subject~\cite{vidaurre2010towards}. 
Hence, frequent calibration is often required in typical HMI systems.
Toward resolving this issue, subject-invariant methods~\cite{wu2020transfer, demir2021autobayes, ozdenizci2019adversarial,  ozdenizci2019transfer,  han2020disentangled, han2021universal, smedemark2021autotransfer}, employing domain generalization and transfer learning, have been proposed to reduce user calibration for HMI systems.

In this paper, we introduce an emerging framework ``quantum machine learning (QML)''~\cite{henderson2020quanvolutional, romero2017quantum, rebentrost2014quantum, lloyd2018quantum, dallaire2018quantum, verdon2019quantum, huggins2019towards, cerezo2021cost, wang2021quantumnas, gomez2022towards, havlivcek2019supervised, schuld2020circuit, farhi2018classification, bergholm2018pennylane, matsumine2019channel, koike2020variational, koike2022quantum, liu2022learning, koike2022autoqml, liu2022variational} into biosignal processing  applications for the first time in the literature, envisioning future era of \emph{quantum supremacy}~\cite{arute2019quantum, zhong2020quantum}.
Quantum computers have the potential to realize computationally efficient signal processing compared to traditional digital computers by exploiting quantum mechanisms, e.g., superposition and entanglement, in terms of not only execution time but also energy consumption.
In the past few years, several vendors have successfully manufactured commercial quantum processing units (QPUs). 
For instance, IBM released $127$-qubit QPUs in 2021, and plans to produce $1121$-qubit QPUs by 2023.
It is thus no longer far future when QML will be widely used for real applications.
Recently, hybrid quantum-classical algorithms based on the \emph{variational} principle~\cite{farhi2014quantum, farhi2016quantum, anschuetz2019variational, kandala2017hardware} were proposed to deal with quantum noise.

The main contributions of this paper are summarized below: 
\begin{itemize}
\item We introduce the emerging QML framework for biosignal processing;
\item We propose a hybrid quantum-classic DNN model called \emph{quEEGNet};
\item We demonstrate the proof-of-concept study on QML for various physiological datasets.
\end{itemize}
To the best of our knowledge, this is the very first research on QML applied to HMI and BCI fields.
Although there exist a few literature~\cite{miranda2020interfacing, swanbci} discussing the potential use of quantum computing for BCI, no practical demonstration on QML-assisted HMI systems is found to date.
Note that our QNN is different from a recurrent QNN (RQNN) employing quantum stochastic filtering based on the Schr\"{o}dinger equation~\cite{gandhi2015evaluating, hao2014stochastic, behera2005quantum, gandhi2014eeg}, which is motivated by quantum physics but does not need real QPUs.
In addition, our work is tangential to quantum sensing technologies such as superconducting quantum interference devices (SQUID)~\cite{ramadan2017brain}.

\section{Quantum Artificial Intelligence (QAI) for HMI}

\subsection{Quantum Bit (Qubit)}

\label{sec:pre}
In quantum systems, a {\em{qubit}} is expressed as the following state superposing bases of $|0\rangle$ and $|1\rangle$:
$| \phi \rangle = \alpha_0 |0\rangle + \alpha_1 |1\rangle$,
where $\alpha_1$ and $\alpha_2$ are complex numbers subject to $|\alpha_0|^2+|\alpha_1|^2=1$.
When qubits are measured, the classical bit $0$ or $1$ is observed with a probability of $|\alpha_0|^2$ or $|\alpha_1|^2$, respectively.
The above \textit{ket-notation} corresponds to column-vector operations of the two basis states $|0\rangle=[1,0]^\mathrm{T}$ and $|1\rangle=[0,1]^\mathrm{T}$, whereas
the \textit{bra-notation} is used for row-vector operations corresponds to its Hermitian transpose; i.e., 
$\langle \phi| = |\phi \rangle^\dag = [\alpha^*_0, \alpha_1^*]$. Here, $[\cdot]^\dag$, $[\cdot]^*$ and $[\cdot]^\mathrm{T}$ denote Hermitian transpose, 
complex conjugate and transpose, respectively.
Note that a multi-qubit state is represented by sum of Kronecker products of basis vectors such as $|000\rangle = |0\rangle^{\otimes 3}$.

\subsection{Quantum Gates}
The basic operations on a qubit is defined as a unitary matrix, which is called {\em{gate}}. Some of the most common gates are associated with Pauli matrices:
$\mathbf{I} = \left[\begin{smallmatrix}
                     1 & 0\\
                     0 & 1
                    \end{smallmatrix}\right]$,
$\mathbf{X} = \left[\begin{smallmatrix}
                     0 & 1\\
                     1 & 0
                    \end{smallmatrix}\right]$,
$\mathbf{Y} = \left[\begin{smallmatrix}
                     0 & -\jmath\\
                     \jmath & 0
                    \end{smallmatrix}\right]$, and
$\mathbf{Z} = \left[\begin{smallmatrix}
                     1 & 0\\
                     0 & -1
                    \end{smallmatrix}\right]$,
where $\jmath$ is the imaginary unit satisfying $\jmath^2=-1$.
The X gate is bit-flip (i.e., NOT operation), Z gate is phase-flip, and Y gate flips both bit and phase.
The Hadamard (H) gate is used to generate a superposition state $|+\rangle = \tfrac{1}{\sqrt{2}} |0\rangle + \tfrac{1}{\sqrt{2}} |1\rangle$:
$\mathbf{H} = \tfrac{1}{\sqrt{2}}
\left[
\begin{smallmatrix}
1 & 1 \\
1 & -1
\end{smallmatrix}
       \right]$.
A controlled-NOT (CNOT or CX) gate is a multi-qubit gate that flips the target qubit if and only if the control qubit is $| 1 \rangle$.

\subsection{Quantum Machine Learning (QML)}

A number of modern DNN methods have been already migrated into the quantum domain, e.g., convolutional layers~\cite{henderson2020quanvolutional}, autoencoders~\cite{romero2017quantum}, graph neural networks~\cite{verdon2019quantum}, and generative adversarial networks~\cite{lloyd2018quantum, dallaire2018quantum}.
Interestingly, the number of QML articles has been exponentially increasing at the same rate of DNN articles, doubling every year, but just $6$ years behind~\cite{koike2022quantum}.
It suggests that QML will be potentially used in numerous communities in a couple of years.
In fact, real QPUs are readily accessible through a cloud quantum server such as IBM QX and Amazon braket.

In analogy with DNN, it was proved that QNN holds the universal approximation property~\cite{perez2020data}.
Accordingly, increasing the number of qubits and quantum layers may enjoy state-of-the-art DNN performance.
In addition, quantum circuits are analytically differentiable~\cite{schuld2019evaluating}, enabling stochastic gradient optimization of QNN.
Nevertheless, QNN often suffers from a vanishing gradient issue called the barren plateau~\cite{mcclean2018barren}.
To mitigate the issue, a simplified 2-design (S2D) ansatz~\cite{cerezo2021cost} was proposed to realize shallow entanglers for arbitrary unitary approximation.
It is highly expected that quantum computers would offer breakthroughs in a wide range of fields. 

\begin{figure}[t]
 \centering 
 \includegraphics[width=\linewidth]
 {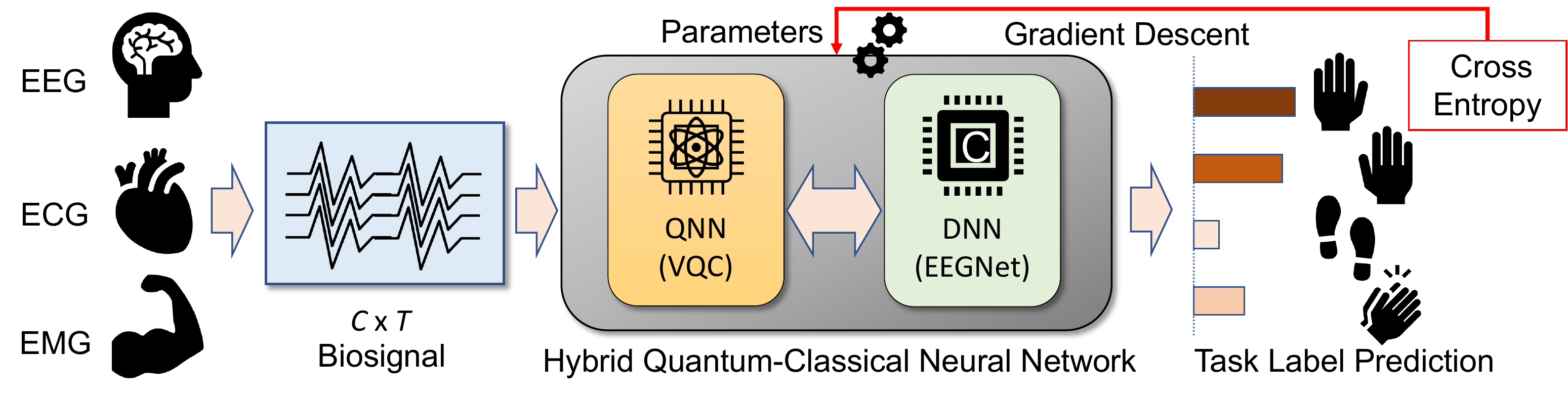}
 \caption{Hybrid quantum-classical neural networks for biosignal processing.}
 \label{fig:queeg}
\end{figure}

\begin{figure}[t]
 \centering 
 \includegraphics[width=\linewidth]
 {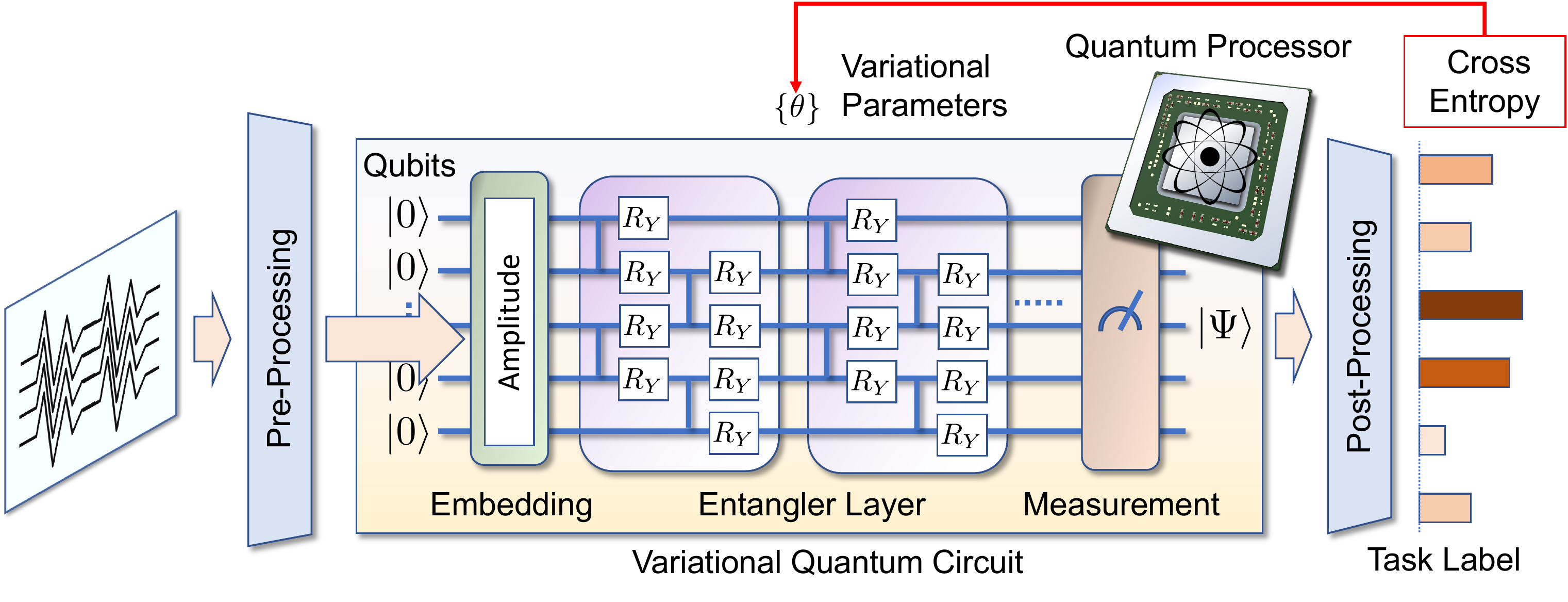}
 \caption{Variational QNN for HMI systems.}
 \label{fig:qnn}
\end{figure}

\subsection{Quantum Neural Network (QNN) for HMI}

Fig.~\ref{fig:queeg} shows an HMI system employing quantum-classical neural network model for biosignal processing.
The system feeds biological waveform arrays to predict a task label through a neural network, which integrates a QNN model with a classical DNN model such as EEGNet~\cite{lawhern2018eegnet}.
The variational parameters for QNN and other trainable parameters for DNN are jointly optimized by stochastic gradient methods to minimize a loss function in an iterative manner.

Fig.~\ref{fig:qnn} depicts an exemplar QNN model based on VQC employing the S2D ansatz~\cite{cerezo2021cost}, which consists of Pauli-Y rotations and staggered controlled-Z entanglers, to evolve the quantum states.
This ansatz is a simplified variant of a $2$-design whose statistical properties are identical to ensemble random unitaries with respect to the Haar measure up to the first $2$ moments.
For an $n$-qubit variational quantum circuit, there are $2(n-1)L$ variational parameters $\{\theta\}$ over an $L$-layer S2D ansatz.

To feed multi-dimensional data, an input layer based on batch normalization is used to initialize the quantum state through the use of an amplitude embedding, which enables encoding up to $2^n-1$ values for $n$-qubit QPUs.
The multi-label task prediction is provided by quantum measurements in the Hamiltonian observable of Pauli-Z operations, followed by a post-processing layer to align the dimension.
The variational parameters as well as input/output layers are optimized by a gradient method to minimize the softmax cross entropy loss.
While QNN is not necessarily better than DNN in prediction accuracy, it can be computationally efficient by manipulating exponentially many quantum states in parallel with a small number of quantum gates.

We integrate the QNN with EEGNet, where the QNN performs as feature extraction and EEGNet works as the post-processing layers.
Note that various other different combinations are possible, e.g., two individual VQC layers for temporal and spatial convolutions; VQC in recurrent networks.
We refer to all such hybrid QNN+DNN concepts (not specific architectures) suited for biological analysis as a \emph{quantum EEGNet (quEEGNet)} by convention.

\section{Experimental Evaluation}

\subsection{Datasets}

We use publicly available physiological datasets, summarized in Table~\ref{tab:data}.
These cover a wide variety of data size, dimensionality, and subject scale as well as sensor modalities, including EEG, EMG, and electrocorticography (ECoG). 

\begin{table}[t]
\centering
\caption{Parameters of Public Dataset Under Investigation}
\label{tab:data}
\scriptsize
\begin{tabular}{c ccc ccc}
\hline
Dataset & Modality & Dimension & Subjects & Classes & Samples \\

\hline

Stress\cite{birjandtalab2016non} & Temp. etc. & $7\times 1$ & $20$ & $4$ & $24{,}000$ \\

RSVP\cite{orhan2012rsvp} & EEG & $16\times 128$ & $10$ & $4$ & $41{,}400$ \\

MI\cite{goldberger2000physiobank} & EEG & $64\times 480$ & $106$ & $4$ & $9{,}540$ \\

ErrP\cite{margaux2012objective} & EEG & $56\times 250$ & $27$ & $2$ & $9{,}180$ \\

Faces Basic\cite{miller2015physiology} & ECoG & $31\times 400$ & $14$ & $2$ & $4{,}100$ \\

Faces Noisy\cite{miller2016spontaneous} & ECoG & $39\times 400$ & $7$ & $2$ & $2{,}100$ \\

ASL\cite{gunay2019transfer} & EMG & $16\times 50$ & $5$ & $33$ & $9{,}900$  \\

\hline
\end{tabular}
\end{table}

\subsubsection{Stress}
    A physiological dataset for neurological stress level\cite{birjandtalab2016non}.\footnote{Stress dataset: \url{https://physionet.org/content/noneeg/1.0.0/}} 
    It consists of multi-modal biosignals for $4$ discrete stress states (physical/cognitive/emotional stresses and relaxation) from $20$ healthy subjects. 
    The data collection consisted of $7$ dimensions of multi-model sensing, i.e., electrodermal activity, temperature, three-dimensional acceleration, heart rate, and arterial oxygen level. 
    A task of $5$ minutes long ($T = 300$ time samples with $1$~Hz down-sampling) was executed for a total of $4$ trials per state.
    
\subsubsection{RSVP}
    An EEG-based typing interface using a rapid serial visual presentation (RSVP) paradigm\cite{orhan2012rsvp}.\footnote{RSVP dataset: \url{http://hdl.handle.net/2047/D20294523}} 
    $10$ healthy subjects participated in the experiments at three sessions performed on different days. 
    The dataset consists of $16$-channel EEG data for $T=128$ samples, which were collected by a g.USBamp amplifier with active electrodes during keyboard operations, for 
    $4$ labels of emotion elicitation, resting-state, or motor imagery/execution task.
    
\subsubsection{MI}
    The PhysioNet EEG motor imagery (MI) dataset\cite{goldberger2000physiobank}.\footnote{MI dataset: \url{https://physionet.org/physiobank/database/eegmmidb/}} 
    Excluding irregular timestamps, the dataset consists of $106$ subjects' EEG data. 
    The subjects were instructed to perform cue-based motor execution/imagery tasks while $64$ channels were recorded at a sampling rate of $160$~Hz. 
    We use the EEG data for three seconds of post-cue interval data ($T=480$ time samples).
    The subject performed $4$-class tasks: right hand motor imagery; left hand motor imagery; both hands motor imagery; or both feet motor imagery. 

\subsubsection{ErrP}
    An error-related potential (ErrP) EEG dataset\cite{margaux2012objective}.\footnote{ErrP dataset: \url{https://www.kaggle.com/c/inria-bci-challenge/}}
    The dataset consists of EEG data recorded from $16$ healthy subjects in an offline P300 spelling task, where visual feedback of the inferred letter is provided at the end of each trial for $1.3$~seconds to monitor evoked brain responses for erroneous decisions made by the system. 
    EEG data were recorded from $56$ channels for epoched $1.25$~seconds at a sampling rate of $200$~Hz ($T=250$).
    Across $5$ recording sessions, each subject performed a total of $340$ trials. 
    
\subsubsection{Faces Basic}
An implanted ECoG array dataset for visual stimulus experiments\cite{miller2015physiology, miller2016spontaneous}.\footnote{Faces dataset: \url{https://exhibits.stanford.edu/data/catalog/zk881ps0522}}
    ECoG arrays were implanted on the subtemporal cortical surface of $14$ epilepsy patients. 
    $2$ classes of grayscale images, either faces or houses, were displayed rapidly in random sequence for $400$~ms.
    The ECoG potentials were measured with respect to a scalp reference and ground, at a sampling rate of $1000$~Hz.
    Subjects performed a basic face-vs-house discrimination task. 
    We use the first $31$ channels to analyze for $T=400$.

\subsubsection{Faces Noisy}
The implanted ECoG arrays dataset for visual stimulus experiments\cite{miller2015physiology, miller2017face}.
    The experiment is similar to Faces Basic dataset, while pictures of faces and houses are randomly scrambled.
    There are $7$ subjects with $39$ channels. 
    Refer ethics statement to reuse the dataset.

\subsubsection{ASL}
An EMG dataset for finger gesture identification for American Sign Language (ASL)\cite{gunay2019transfer}.\footnote{ASL Dataset: \url{http://hdl.handle.net/2047/D20294523}}
    $5$ healthy, right-handed, subjects participated in experiments with surface EMG (Delsys Trigno) recorded at $2$~kHz from $16$ lower-arm muscles.
    Subjects shaped their right hand into an ASL posture presented on a video screen ($33$ postures, $3$ trials per posture).
    Dynamic letters `J' and `Z' were omitted, along with the number `0', which is confusing with the letter `O'.
    The participants were given $2$~seconds to form the posture and $6$~seconds to maintain.
    The signal is decimated to be $T=50$.

\subsection{Model Implementation}

We use PennyLane and PyTorch libraries to train quEEGNet.
The trainable parameters are optimized by the adaptive momentum (Adam) with a learning rate of $0.1$ for $50$ epochs with a batch size of $128$.

\subsection{Performance Results}

Table~\ref{tab:result} shows the performance comparison between EEGNet and quEEGNet.
It was verified that the hybrid quantum-classical model outperforms classical neural networks for all of the physiological datasets.
Since we have not explored different variants of quantum ansatz yet, it is expected that the performance can be further improved via AutoQML~\cite{koike2022autoqml}.

\begin{table}[t]
\centering
\caption{Performance Results in Test Accuracy (\%)}
\label{tab:result}
\scriptsize
\begin{tabular}{c cc}
\hline
Dataset & EEGNet & quEEGNet \\

\hline

Stress & $85.87$ & $87.23$ \\

RSVP & $93.73$ & $95.12$ \\

MI & $59.61$ & $60.22$ \\

ErrP & $74.36$ & $75.92$ \\

Faces Basic & $63.30$ & $64.92$ \\

Faces Noisy & $75.94$ & $78.01$ \\

ASL & $23.64$ & $25.16$  \\

\hline
\end{tabular}
\end{table}

\section{Conclusions}

We proposed an emerging QML framework for HMI/BCI systems, considering
the recent rapid advancement of quantum technology.
Our hybrid quantum-classical neural network was demonstrated to achieve the state-of-the-art performance for various physiological datasets.
As the application of QML to HMI/BCI fields is still at a proof-of-concept phase, there remain many open problems to explore for future work.

\bibliographystyle{IEEEtran}
\bibliography{ref, quantum}

\begin{thebibliography}{10}
\providecommand{\url}[1]{#1}
\csname url@samestyle\endcsname
\providecommand{\newblock}{\relax}
\providecommand{\bibinfo}[2]{#2}
\providecommand{\BIBentrySTDinterwordspacing}{\spaceskip=0pt\relax}
\providecommand{\BIBentryALTinterwordstretchfactor}{4}
\providecommand{\BIBentryALTinterwordspacing}{\spaceskip=\fontdimen2\font plus
\BIBentryALTinterwordstretchfactor\fontdimen3\font minus
  \fontdimen4\font\relax}
\providecommand{\BIBforeignlanguage}[2]{{%
\expandafter\ifx\csname l@#1\endcsname\relax
\typeout{** WARNING: IEEEtran.bst: No hyphenation pattern has been}%
\typeout{** loaded for the language `#1'. Using the pattern for}%
\typeout{** the default language instead.}%
\else
\language=\csname l@#1\endcsname
\fi
#2}}
\providecommand{\BIBdecl}{\relax}
\BIBdecl

\bibitem{faust2018deep}
O.~Faust, Y.~Hagiwara, T.~J. Hong, O.~S. Lih, and U.~R. Acharya, ``Deep
  learning for healthcare applications based on physiological signals: A
  review,'' \emph{Computer methods and programs in biomedicine}, vol. 161, pp.
  1--13, 2018.

\bibitem{lawhern2018eegnet}
V.~J. Lawhern, A.~J. Solon, N.~R. Waytowich, S.~M. Gordon, C.~P. Hung, and
  B.~J. Lance, ``{EEGNet}: a compact convolutional neural network for
  {EEG}-based brain--computer interfaces,'' \emph{Journal of neural
  engineering}, vol.~15, no.~5, p. 056013, 2018.

\bibitem{atzori2016deep}
M.~Atzori, M.~Cognolato, and H.~M{\"u}ller, ``Deep learning with convolutional
  neural networks applied to electromyography data: A resource for the
  classification of movements for prosthetic hands,'' \emph{Frontiers in
  neurorobotics}, vol.~10, p.~9, 2016.

\bibitem{vidaurre2010towards}
C.~Vidaurre and B.~Blankertz, ``Towards a cure for {BCI} illiteracy,''
  \emph{Brain topography}, vol.~23, no.~2, pp. 194--198, 2010.

\bibitem{wu2020transfer}
D.~Wu, Y.~Xu, and B.-L. Lu, ``Transfer learning for {EEG}-based brain--computer
  interfaces: A review of progress made since 2016,'' \emph{IEEE Transactions
  on Cognitive and Developmental Systems}, vol.~14, no.~1, pp. 4--19, 2020.

\bibitem{demir2021autobayes}
A.~Demir, T.~Koike-Akino, Y.~Wang, and D.~Erdogmus, ``{AutoBayes}: Automated
  {Bayesian} graph exploration for nuisance-robust inference,'' \emph{IEEE
  Access}, vol.~9, pp. 39\,955--39\,972, 2021.

\bibitem{ozdenizci2019adversarial}
O.~{\"O}zdenizci, Y.~Wang, T.~Koike-Akino, and D.~Erdo{\u{g}}mu{\c{s}},
  ``Adversarial deep learning in {EEG} biometrics,'' \emph{IEEE signal
  processing letters}, vol.~26, no.~5, pp. 710--714, 2019.

\bibitem{ozdenizci2019transfer}
------, ``Transfer learning in brain-computer interfaces with adversarial
  variational autoencoders,'' in \emph{2019 9th International IEEE/EMBS
  Conference on Neural Engineering (NER)}.\hskip 1em plus 0.5em minus
  0.4em\relax IEEE, 2019, pp. 207--210.

\bibitem{han2020disentangled}
M.~Han, O.~{\"O}zdenizci, Y.~Wang, T.~Koike-Akino, and D.~Erdo{\u{g}}mu{\c{s}},
  ``Disentangled adversarial autoencoder for subject-invariant physiological
  feature extraction,'' \emph{IEEE signal processing letters}, vol.~27, pp.
  1565--1569, 2020.

\bibitem{han2021universal}
M.~Han, O.~{\"O}zdenizci, T.~Koike-Akino, Y.~Wang, and D.~Erdo{\u{g}}mu{\c{s}},
  ``Universal physiological representation learning with soft-disentangled
  rateless autoencoders,'' \emph{IEEE Journal of Biomedical and Health
  Informatics}, vol.~25, no.~8, pp. 2928--2937, 2021.

\bibitem{smedemark2021autotransfer}
N.~Smedemark-Margulies, Y.~Wang, T.~Koike-Akino, and D.~Erdogmus,
  ``{AutoTransfer}: Subject transfer learning with censored representations on
  biosignals data,'' \emph{arXiv preprint arXiv:2112.09796}, 2021.

\bibitem{henderson2020quanvolutional}
M.~Henderson, S.~Shakya, S.~Pradhan, and T.~Cook, ``Quanvolutional neural
  networks: powering image recognition with quantum circuits,'' \emph{Quantum
  Machine Intelligence}, vol.~2, no.~1, pp. 1--9, 2020.

\bibitem{romero2017quantum}
J.~Romero, J.~P. Olson, and A.~Aspuru-Guzik, ``Quantum autoencoders for
  efficient compression of quantum data,'' \emph{Quantum Science and
  Technology}, vol.~2, no.~4, p. 045001, 2017.

\bibitem{rebentrost2014quantum}
P.~Rebentrost, M.~Mohseni, and S.~Lloyd, ``Quantum support vector machine for
  big data classification,'' \emph{Physical review letters}, vol. 113, no.~13,
  p. 130503, 2014.

\bibitem{lloyd2018quantum}
S.~Lloyd and C.~Weedbrook, ``Quantum generative adversarial learning,''
  \emph{Physical review letters}, vol. 121, no.~4, p. 040502, 2018.

\bibitem{dallaire2018quantum}
P.-L. Dallaire-Demers and N.~Killoran, ``Quantum generative adversarial
  networks,'' \emph{Physical Review A}, vol.~98, no.~1, p. 012324, 2018.

\bibitem{verdon2019quantum}
G.~Verdon, T.~McCourt, E.~Luzhnica, V.~Singh, S.~Leichenauer, and J.~Hidary,
  ``Quantum graph neural networks,'' \emph{arXiv preprint arXiv:1909.12264},
  2019.

\bibitem{huggins2019towards}
W.~Huggins, P.~Patil, B.~Mitchell, K.~B. Whaley, and E.~M. Stoudenmire,
  ``Towards quantum machine learning with tensor networks,'' \emph{Quantum
  Science and technology}, vol.~4, no.~2, p. 024001, 2019.

\bibitem{cerezo2021cost}
M.~Cerezo, A.~Sone, T.~Volkoff, L.~Cincio, and P.~J. Coles, ``Cost function
  dependent barren plateaus in shallow parametrized quantum circuits,''
  \emph{Nature communications}, vol.~12, no.~1, pp. 1--12, 2021.

\bibitem{wang2021quantumnas}
H.~Wang, Y.~Ding, J.~Gu, Y.~Lin, D.~Z. Pan, F.~T. Chong, and S.~Han,
  ``{QuantumNAS}: Noise-adaptive search for robust quantum circuits,''
  \emph{arXiv preprint arXiv:2107.10845}, 2021.

\bibitem{gomez2022towards}
R.~B. G{\'o}mez, C.~O'Meara, G.~Cortiana, C.~B. Mendl, and
  J.~Bernab{\'e}-Moreno, ``Towards autoqml: A cloud-based automated circuit
  architecture search framework,'' \emph{arXiv preprint arXiv:2202.08024},
  2022.

\bibitem{havlivcek2019supervised}
V.~Havl{\'\i}{\v{c}}ek, A.~D. C{\'o}rcoles, K.~Temme, A.~W. Harrow, A.~Kandala,
  J.~M. Chow, and J.~M. Gambetta, ``Supervised learning with quantum-enhanced
  feature spaces,'' \emph{Nature}, vol. 567, no. 7747, pp. 209--212, 2019.

\bibitem{schuld2020circuit}
M.~Schuld, A.~Bocharov, K.~M. Svore, and N.~Wiebe, ``Circuit-centric quantum
  classifiers,'' \emph{Physical Review A}, vol. 101, no.~3, p. 032308, 2020.

\bibitem{farhi2018classification}
E.~Farhi and H.~Neven, ``Classification with quantum neural networks on near
  term processors,'' \emph{arXiv preprint arXiv:1802.06002}, 2018.

\bibitem{bergholm2018pennylane}
V.~Bergholm, J.~Izaac, M.~Schuld, C.~Gogolin, M.~S. Alam, S.~Ahmed, J.~M.
  Arrazola, C.~Blank, A.~Delgado, S.~Jahangiri \emph{et~al.}, ``Pennylane:
  Automatic differentiation of hybrid quantum-classical computations,''
  \emph{arXiv preprint arXiv:1811.04968}, 2018.

\bibitem{matsumine2019channel}
T.~Matsumine, T.~Koike-Akino, and Y.~Wang, ``Channel decoding with quantum
  approximate optimization algorithm,'' in \emph{2019 IEEE International
  Symposium on Information Theory (ISIT)}.\hskip 1em plus 0.5em minus
  0.4em\relax IEEE, 2019, pp. 2574--2578.

\bibitem{koike2020variational}
T.~Koike-Akino, T.~Matsumine, Y.~Wang, D.~S. Millar, K.~Kojima, and K.~Parsons,
  ``Variational quantum demodulation for coherent optical multi-dimensional
  {QAM},'' in \emph{2020 Optical Fiber Communications Conference and Exhibition
  (OFC)}.\hskip 1em plus 0.5em minus 0.4em\relax OSA, 2020, pp. 1--3.

\bibitem{koike2022quantum}
T.~Koike-Akino, P.~Wang, and Y.~Wang, ``Quantum transfer learning for {Wi-Fi}
  sensing,'' \emph{arXiv preprint arXiv:2205.08590}, 2022.

\bibitem{liu2022learning}
B.~Liu, T.~Koike-Akino, Y.~Wang, and K.~Parsons, ``Learning to learn quantum
  turbo detection,'' \emph{arXiv preprint arXiv:2205.08611}, 2022.

\bibitem{koike2022autoqml}
T.~Koike-Akino, P.~Wang, and Y.~Wang, ``{AutoQML}: Automated quantum machine
  learning for {Wi-Fi} integrated sensing and communications,'' \emph{arXiv
  preprint arXiv:2205.09115}, 2022.

\bibitem{liu2022variational}
B.~Liu, T.~Koike-Akino, Y.~Wang, and K.~Parsons, ``Variational quantum
  compressed sensing for joint user and channel state acquisition in grant-free
  device access systems,'' \emph{arXiv preprint arXiv:2205.08603}, 2022.

\bibitem{arute2019quantum}
F.~Arute, K.~Arya, R.~Babbush, D.~Bacon, J.~C. Bardin, R.~Barends, R.~Biswas,
  S.~Boixo, F.~G. Brandao, D.~A. Buell \emph{et~al.}, ``Quantum supremacy using
  a programmable superconducting processor,'' \emph{Nature}, vol. 574, no.
  7779, pp. 505--510, 2019.

\bibitem{zhong2020quantum}
H.-S. Zhong, H.~Wang, Y.-H. Deng, M.-C. Chen, L.-C. Peng, Y.-H. Luo, J.~Qin,
  D.~Wu, X.~Ding, Y.~Hu \emph{et~al.}, ``Quantum computational advantage using
  photons,'' \emph{Science}, vol. 370, no. 6523, pp. 1460--1463, 2020.

\bibitem{farhi2014quantum}
E.~Farhi, J.~Goldstone, and S.~Gutmann, ``A quantum approximate optimization
  algorithm,'' \emph{arXiv preprint arXiv:1411.4028}, 2014.

\bibitem{farhi2016quantum}
E.~Farhi and A.~W. Harrow, ``Quantum supremacy through the quantum approximate
  optimization algorithm,'' \emph{arXiv preprint arXiv:1602.07674}, 2016.

\bibitem{anschuetz2019variational}
E.~Anschuetz, J.~Olson, A.~Aspuru-Guzik, and Y.~Cao, ``Variational quantum
  factoring,'' in \emph{International Workshop on Quantum Technology and
  Optimization Problems}.\hskip 1em plus 0.5em minus 0.4em\relax Springer,
  2019, pp. 74--85.

\bibitem{kandala2017hardware}
A.~Kandala, A.~Mezzacapo, K.~Temme, M.~Takita, M.~Brink, J.~M. Chow, and J.~M.
  Gambetta, ``Hardware-efficient variational quantum eigensolver for small
  molecules and quantum magnets,'' \emph{Nature}, vol. 549, no. 7671, pp.
  242--246, 2017.

\bibitem{miranda2020interfacing}
E.~R. Miranda, ``On interfacing the brain with quantum computers: An approach
  to listen to the logic of the mind,'' \emph{arXiv preprint arXiv:2101.03887},
  2020.

\bibitem{swanbci}
M.~Swan, ``{BCI} quantum computing {IPLD} for brain,'' \emph{ResearchGate
  preprint:342184271}, Jun. 2020.

\bibitem{gandhi2015evaluating}
V.~Gandhi, G.~Prasad, D.~Coyle, L.~Behera, and T.~M. McGinnity, ``Evaluating
  quantum neural network filtered motor imagery brain-computer interface using
  multiple classification techniques,'' \emph{Neurocomputing}, vol. 170, pp.
  161--167, 2015.

\bibitem{hao2014stochastic}
W.~Hao-Han, J.~Fu-Jiang, L.~Lian-You, and W.~Liang, ``A stochastic filtering
  algorithm using {S}chr{\"o}dinger equation,'' \emph{Acta Automatica Sinica},
  vol.~40, no.~10, pp. 2370--2376, 2014.

\bibitem{behera2005quantum}
L.~Behera and I.~Kar, ``Quantum stochastic filtering,'' in \emph{2005 IEEE
  International Conference on Systems, Man and Cybernetics}, vol.~3.\hskip 1em
  plus 0.5em minus 0.4em\relax IEEE, 2005, pp. 2161--2167.

\bibitem{gandhi2014eeg}
V.~Gandhi, G.~Prasad, D.~Coyle, L.~Behera, and T.~M. McGinnity, ``{EEG}-based
  mobile robot control through an adaptive brain--robot interface,'' \emph{IEEE
  Transactions on Systems, Man, and Cybernetics: Systems}, vol.~44, no.~9, pp.
  1278--1285, 2014.

\bibitem{ramadan2017brain}
R.~A. Ramadan and A.~V. Vasilakos, ``Brain computer interface: control signals
  review,'' \emph{Neurocomputing}, vol. 223, pp. 26--44, 2017.

\bibitem{perez2020data}
A.~P{\'e}rez-Salinas, A.~Cervera-Lierta, E.~Gil-Fuster, and J.~I. Latorre,
  ``Data re-uploading for a universal quantum classifier,'' \emph{Quantum},
  vol.~4, p. 226, 2020.

\bibitem{schuld2019evaluating}
M.~Schuld, V.~Bergholm, C.~Gogolin, J.~Izaac, and N.~Killoran, ``Evaluating
  analytic gradients on quantum hardware,'' \emph{Physical Review A}, vol.~99,
  no.~3, p. 032331, 2019.

\bibitem{mcclean2018barren}
J.~R. McClean, S.~Boixo, V.~N. Smelyanskiy, R.~Babbush, and H.~Neven, ``Barren
  plateaus in quantum neural network training landscapes,'' \emph{Nature
  communications}, vol.~9, no.~1, pp. 1--6, 2018.

\bibitem{birjandtalab2016non}
J.~Birjandtalab, D.~Cogan, M.~B. Pouyan, and M.~Nourani, ``A non-{EEG}
  biosignals dataset for assessment and visualization of neurological status,''
  in \emph{2016 IEEE International Workshop on Signal Processing Systems
  (SiPS)}.\hskip 1em plus 0.5em minus 0.4em\relax IEEE, 2016, pp. 110--114.

\bibitem{orhan2012rsvp}
U.~Orhan, K.~E. Hild, D.~Erdo{\u{g}}mu{\c{s}}, B.~Roark, B.~Oken, and
  M.~Fried-Oken, ``{RSVP} keyboard: An {EEG} based typing interface,'' in
  \emph{2012 IEEE International Conference on Acoustics, Speech and Signal
  Processing (ICASSP)}.\hskip 1em plus 0.5em minus 0.4em\relax IEEE, 2012, pp.
  645--648.

\bibitem{goldberger2000physiobank}
A.~L. Goldberger, L.~A. Amaral, L.~Glass, J.~M. Hausdorff, P.~C. Ivanov, R.~G.
  Mark, J.~E. Mietus, G.~B. Moody, C.-K. Peng, and H.~E. Stanley, ``Physiobank,
  physiotoolkit, and physionet: components of a new research resource for
  complex physiologic signals,'' \emph{circulation}, vol. 101, no.~23, pp.
  e215--e220, 2000.

\bibitem{margaux2012objective}
P.~Margaux, M.~Emmanuel, D.~S{\'e}bastien, B.~Olivier, and M.~J{\'e}r{\'e}mie,
  ``Objective and subjective evaluation of online error correction during
  {P300}-based spelling,'' \emph{Advances in Human-Computer Interaction}, vol.
  2012, 2012.

\bibitem{miller2015physiology}
K.~J. Miller, D.~Hermes, N.~Witthoft, R.~P. Rao, and J.~G. Ojemann, ``The
  physiology of perception in human temporal lobe is specialized for contextual
  novelty,'' \emph{Journal of neurophysiology}, vol. 114, no.~1, pp. 256--263,
  2015.

\bibitem{miller2016spontaneous}
K.~J. Miller, G.~Schalk, D.~Hermes, J.~G. Ojemann, and R.~P. Rao, ``Spontaneous
  decoding of the timing and content of human object perception from cortical
  surface recordings reveals complementary information in the event-related
  potential and broadband spectral change,'' \emph{PLoS computational biology},
  vol.~12, no.~1, 2016.

\bibitem{gunay2019transfer}
S.~Y. G{\"u}nay, M.~Yarossi, D.~H. Brooks, E.~Tunik, and
  D.~Erdo{\u{g}}mu{\c{s}}, ``Transfer learning using low-dimensional subspaces
  for {EMG}-based classification of hand posture,'' in \emph{2019 9th
  International IEEE/EMBS Conference on Neural Engineering (NER)}.\hskip 1em
  plus 0.5em minus 0.4em\relax IEEE, 2019, pp. 1097--1100.

\bibitem{miller2017face}
K.~J. Miller, D.~Hermes, F.~Pestilli, G.~S. Wig, and J.~G. Ojemann, ``Face
  percept formation in human ventral temporal cortex,'' \emph{Journal of
  neurophysiology}, vol. 118, no.~5, pp. 2614--2627, 2017.

\end{thebibliography}
\end{document}